\documentclass[12pt]{article}
\usepackage{pic03}
\usepackage{hyperref}
\usepackage{url}
\usepackage{graphicx}

\begin{document}

\title{\bf HEAVY QUARKONIUM HYPERFINE STRUCTURE AND LEPTONIC DECAYS}
\author{D. Ebert$^a$, R. N. Faustov$^{a,b}$, V. O. Galkin$^{a,b}$\\
{\em $^a$Institut f\"ur Physik,
  Humboldt--Universit\"at zu Berlin, Berlin, 
  Germany}\\
{\em $\!\!\!\!^b$Russian Academy of Sciences, Scientific Council for
Cybernetics, Moscow, Russia} }
\maketitle

\baselineskip=14.5pt
\begin{abstract}
The relation between the hyperfine structure and leptonic decay rates
of heavy quarkonia is considered with the account of relativistic
and  radiative corrections. The calculated decay rates agree well with
the available experimental data, while the predicted $\eta_c(2S)$ mass
is significantly smaller than the value measured recently by the Belle
and BaBar Collaborations. 

\end{abstract}

\baselineskip=17pt

\noindent
Recent observation of the charmonium $2^1\!S_0$ state $\eta_c(2S)$ by
the Belle and BaBar Collaborations yielded the following values for
its mass: 
\begin{equation}
  \label{eq:bv}
  M(\eta_c(2S))=\left\{
      \begin{array}{l}
3.654(14)\ \textrm{GeV  (see Ref. \cite{belle1})} \cr
3.622(12)\ \textrm{GeV  (see Ref. \cite{belle2})}\cr
3.632(6)\ \ \, \textrm{GeV   (see Ref. \cite{babar})}
      \end{array}\right. .
\end{equation}
These values are significantly larger than most predictions of the
constituent quark models \cite{efg,eq} and the previous (unconfirmed)
experimental value \cite{cbc} $M[\eta_c(2S)]=3.594(5)$~GeV. The resulting $2S$
hyperfine splitting (HFS) would be about 2--4 times smaller than the
$1S$ HFS in charmonium which is quite unexpected \cite{eq,bb}. The HFS
is closely connected with leptonic decay rates and vector decay constants of
heavy quarkonia. This connection for charmonium was discussed at
length in the recent paper \cite{bb}. Here we extend this discussion in
order to include the relativistic corrections for the vector
constants. We considered these corrections for heavy-light
($B$ and $D$) mesons in our recent paper \cite{fconst}.
The vector decay constant $f_V$ is defined by
$  \left<0|\bar Q\gamma^\mu Q|V({\bf K},\varepsilon)\right>=f_V
  M\varepsilon^\mu,$
where ${\bf K}$ is the quarkonium momentum, $\varepsilon$ and $M$ are
the polarization vector and mass of the quarkonium. The relativistic
expression for $f_V$ can be obtained from Eq.~(3) of
Ref.~\cite{fconst} by putting $m_q=m_Q=m$   
\begin{equation}
  \label{eq:fv}
 f_V=\sqrt{\frac{12}{M}}\int\frac{d^3p}{(2\pi)^3}
 \left\{1-\frac{\epsilon(p)-m}{3\epsilon(p)}\right\}\Phi_V(p), 
\end{equation}
where $\epsilon(p)=\sqrt{{\bf p}^2+m^2}$ and $\Phi_V(p)$ is the vector
quarkonium wave function in the momentum space.
In the nonrelativistic limit $p^2/m^2\to 0$ this expression reduces to the
well-known formula
$  f_V^{\rm NR}=\sqrt{{12}/{M}}\left|\Psi_V(0)\right|$,
where $\Psi_V(0)$ is the wave function at the origin $r=0$.
%
The leptonic decay rate for zero lepton mass is given by 
\begin{equation}\label{eq:gamma}
 \Gamma_0(V\to e^+e^-)=\frac{4\pi\alpha^2e_Q^2}{3M}f_V^2,
\end{equation}
where $\alpha$ is the QED fine structure constant and $e_Q$ is the
quark charge in units of the elementary electric charge.  In the
nonrelativistic limit we obtain the widely-used
relation
$  \Gamma^{\rm{NR}}_0=({16\pi\alpha^2e_Q^2}/{M^2})
  \left|\Psi_V(0)\right|^2. $
The one-loop QCD corrections modify Eq.~(\ref{eq:gamma}) in the
following way \cite{bgkk}
$  \Gamma=\Gamma_0\left(1-\frac{16}{3\pi}\alpha_s\right),$
where we take the QCD coupling constant $\alpha_s$ according to
PDG \cite{pdg} equal to
$\alpha_s(m_c)=0.26$,
$ \alpha_s(m_b)=0.18.$

\begin{table}[htb]
\centering
\caption{\it $1S$ and $2S$ heavy quarkonium masses (in GeV) and HFS
  $\Delta M$ (in MeV).} 
 \vskip 0.09 in
{\begin{tabular}{@{}ccccccc@{}} \hline
States& $1^1\!S_0$&$1^3\!S_1$ &$\Delta M^{\rm{HFS}}_{1S}$&
$2^1\!S_0$&$2^3\!S_1$ &$\Delta M^{\rm{HFS}}_{2S}$ \\\hline
$c\bar c$& $\eta_c(1S)$& $J/\psi(1S)$& &$\eta_c(2S)$ &$\psi(2S)$& \\
theory \cite{efg}&2.979& 3.096& 117& 3.588& 3.686& 98\\
exp. \cite{pdg}&2.9797(15)&3.09687(4)&117(1)  & & 3.68596(9)&\\
exp. \cite{belle1}&2.979(2)&    & 117(1)      &3.654(14)&
&32(14)\\ 
exp. \cite{belle2}&2.979(2)&    & 117(1)      &3.622(12)&
&64(12)\\
exp. \cite{babar}&         &    &        & 3.632(6)& &54(6)\\
exp. \cite{cbc}& &   & & 3.594(5)&  & 92(5)\\
\hline
$b\bar b$&$\eta_b(1S)$& $\Upsilon(1S)$& &$\eta_b(2S)$ &$\Upsilon(2S)$&
\\
theory \cite{efg}& 9.400& 9.460 & 60& 9.993 & 10.023& 30\\
exp. \cite{pdg}& & 9.46030(26)& & &10.02326(31)
\\ \hline
\end{tabular}}
\end{table} 

\noindent
The heavy quarkonium mass spectra including all relativistic $v^2/c^2$
and one-loop radiative corrections were calculated in
Ref.~\cite{efg}. The masses $M$ and HFS $\Delta
M^{\rm{HFS}}_{nS}\equiv M(n^3\!S_1)-M(n^1\!S_0)$ of $nS$  states
($n=1,2$) are presented in Table~1.
Inserting into Eq.~(\ref{eq:fv}) the same wave functions as used in
Ref.~\cite{efg} we can calculate  the leptonic decay rates of
vector quarkonia. The results of these calculations in comparison with
experimental data are shown in Table~2.
Relativistic corrections considerably reduce the
calculated decay rates and bring them in good agreement with
experimental values. Thus we observe the overall selfconsistency
between predictions for HFS and leptonic decay rates of heavy
quarkonia. It means that the new Belle and BaBar data
\cite{belle1,belle2,babar} for the 
$\eta_c(2S)$ mass cannot be easily explained in the framework of
relativistic constituent quark models and, if confirmed, require some
novel ideas and approaches (scale choice \cite{bb} for $\alpha_s$, state
mixing \cite{mr}, etc.). 
This work
was supported in part by the {\it Deutsche
Forschungsgemeinschaft} under contract Eb 139/2-2.

\begin{table}[t]
\centering
\caption{\it Vector decay constants and leptonic decay rates of vector
  quarkonia.} 
\vskip 0.1 in
{\begin{tabular}{@{}ccccc@{}}\hline 
Decay&$f_V$&$\Gamma^{\rm NR}$&$\Gamma$&$\Gamma^{\rm exp}$\\
modes& MeV& keV&  keV& keV\\ \hline
$J/\psi(1S)\to e^+e^-$ &551 & 6.7 & 5.4 &5.26(37)\\
$\psi(2S)\to e^+e^-$ &401  & 3.2 & 2.4 &2.19(15)\\
$\Upsilon(1S)\to e^+e^-$ &839 & 1.4 & 1.3 & 1.32(5)\\
$\Upsilon(2S)\to e^+e^-$ &562 & 0.6 & 0.5 & 0.52(3)
\\ \hline
\end{tabular}}
\end{table}


\end{document}